\documentclass[letter]{aa}
\usepackage[dvips]{graphicx}
\usepackage{txfonts}
\usepackage{natbib}
\bibpunct{(}{)}{;}{a}{}{,}

\begin{document}

\title{RXTE confirmation of the Intermediate Polar status of Swift~J0732.5--1331}

\author{O.W. Butters\inst{1}
\and
  E.J. Barlow\inst{1}
\and  
  A.J. Norton\inst{1}
\and
  K. Mukai\inst{2}
}

\institute{
  Department of Physics and Astronomy, The Open University, Walton
  Hall, Milton Keynes MK7 6AA, UK\\
  \email{o.w.butters@open.ac.uk}
\and
 CRESST and X-ray Astrophysics Laboratory NASA/GSFC, Greenbelt,
        MD 20771, USA {\it and} Department of Physics, University of Maryland,
        Baltimore county, 1000 Hilltop Circle, Baltimore, MD 21250, USA 
}

\authorrunning{Butters et al.}

\date{Accepted 2007 ???;
      Received  2007 ???;
      in original form 2007 ???}

\abstract
{}
{We intend to establish the X-ray properties of Swift J0732.5--1331 and therefore
 confirm its status as an intermediate polar.}
{We analysed 36\,240~s of X-ray data from {\em RXTE}. 
Frequency analysis was used to constrain temporal variations and
spectral analysis used to characterise the emission and
  absorption properties.}
{The X-ray spin period is confirmed to be 512.4(3)~s with a strong first
harmonic. No modulation is detected at the candidate orbital period of
5.6~h, but a coherent modulation is present at the candidate 11.3~h
period. The spectrum is consistent with a 37~keV
bremsstrahlung continuum with an iron line at 6.4~keV absorbed by an
equivalent hydrogen column density of around $10^{22}$ atoms~cm$^{-2}$.}
{Swift J0732--1331 is confirmed to be an intermediate polar.}

\keywords{stars:binary -- stars:novae, cataclysmic variables -- stars:
  individual:Swift~J0732.5--1331 -- X-rays: binaries  }

\maketitle

\section{Introduction to magnetic cataclysmic variables}

Intermediate polars (IPs) are a sub-class of cataclysmic variables (CVs). They
fill the phase space, in terms of magnetic field strength, and spin and
orbital periods, between non-magnetic CVs and the strongly
magnetic synchronously rotating polars. The magnetic field strength is
believed to be in the range of a few MG to tens of MG at the white
dwarf surface. This is large enough to dramatically alter the accretion flow,
yet not large enough to synchronize the spin and orbital periods. This
magnetic field gives rise to the defining characteristic of the
sub-class, that of X-ray variation pulsed at the spin period of the
white dwarf. For an exhaustive review of CVs see e.g. \cite{warner95}.

There are between twenty
six\footnote{http://asd.gsfc.nasa.gov/Koji.Mukai/iphome/iphome.html
as of 23/8/7.} and fifty IPs currently known (depending on the selection criteria
used). The hard X-ray selected object, Swift~J0732.5--1331 (hereafter J0732),
is a suspected IP in need of confirmation. The circumstance of its
discovery makes J0732 similar to the host of candidate IPs that have been
discovered to be powerful emitters of hard X-rays/soft gamma-rays in
the 20--100~keV range in the {\sl INTEGRAL\/}/IBIS survey
\citep{barlow06}. We have embarked on a campaign of pointed {\sl
  RXTE\/} observations of these hard X-ray discovered candidate
IPs. Here we present the first results of our campaign on J0732.

\section{Previous observations of Swift J0732.5-1331}

There is to date no peer-reviewed analysis of J0732 published in the 
literature. There are however several mentions of it in Astronomical 
Telegrams, which we summarise below. These were all published over the 
course of a couple of months in early 2006.

J0732 was first detected by \cite{ajello06} with the {\em Swift} Burst Alert
Telescope and {\em Swift} X-ray telescope (XRT). With the XRT 600 counts in
3,400~s were recorded, coincident with the {\em ROSAT} source 
1RXS~J073237.6--133113. This is also coincident with the 2MASS source
J073237.64--133109.4, a proposed K main-sequence star 400~pc away.
Based on its X-ray luminosity and colours, \cite{ajello06} suggested 
a CV identification for the object.

\cite{masetti06} used the BFOSC instrument on the G.D. Cassini 1.5~m
telescope to obtain the optical spectrum of the counterpart to
J0732. Two objects were found close to the reported position, a normal 
G/K type Galactic star (the 2MASS source) and a fainter one deemed to be 
the true optical counterpart. The spectral signature of the system 
was concluded to be that of an accretion disc in a low mass X-ray binary.

\cite{patterson06} also obtained low resolution spectra, this time
on the MDM 2.4~m telescope, of the 2MASS optical counterpart proposed by
\cite{ajello06} (i.e. the field star), concluding that it was indeed a 
normal G star. In the same telegram \cite{patterson06} also reported 
optical photometry (obtained by the small telescope network of the 
Center for Backyard Astrophysics\footnote{http://cba.phys.columbia.edu/})
which revealed a stable pulsation period of 512.42(3)~s with most of the power 
in the first harmonic. This was deemed to be the spin period of a rotating
white dwarf and \cite{patterson06} consequently suggested an IP classification
for the object. A possible 11.3 hour orbital signal was also suggested, but 
owing to its low amplitude, this required the binary to be close to face on, 
as any variation in brightness due to the projected Roche lobe filling secondary
was small.

\cite{marsh06} subsequently used {\em ULTRACAM} mounted on the William 
Herschel Telescope to observe the optical counterpart of J0732. The spin 
pulsation detected by \cite{patterson06} was seen. The counterpart and 
the non-associated field star were found to be approximately 
$1.8^{\prime \prime}$ apart.

\cite{torres06} performed spectral analysis of the optical counterpart at 
the Mt. Hopkins 1.5~m telescope. Balmer emission lines from H$\alpha$ 
to at least H$\gamma$ were found. This reinforced its classification as a
probable IP. However, despite seeing a variation in the radial velocities of the
various emission lines, they were unable to determine an orbital period.

\cite{wheatley06} later reanalysed the original {\em Swift} data reported in
\cite{ajello06} and found an X-ray pulsation at the proposed spin period. The
modulation was found to be single peaked and only present below
2~keV. Reanalysis of the spectral data suggested a temperature 
typical of intermediate polars ($kT\sim$20~keV). This data set is, however, 
short and suffers from severe aliasing effects.

\cite{thorstensen06} carried out 
time series spectroscopy at the MDM Observatory on the optical counterpart. 
The radial velocities of the H$\alpha$ emission lines were found to vary 
periodically with a period of 0.2335(8)~days (5.60(2)~h), which was interpreted as 
the orbital period of the system. We note this is close to half the photometric 
period suggested by \cite{patterson06}.

Given all this information, J0732 is strongly suspected to be an
intermediate polar, but the only way to confirm this is the
unambiguous detection of pulsed hard X-ray emission at the spin period.

\section{Observations and data reduction}

Data were obtained from the {\em RXTE} satellite \citep{bradt93} with the PCA 
instrument over two consecutive days, from 13th -- 15th July 2007. The total time 
on target was 36\,240~s, comprising fourteen approximately equal segments of one 
satellite orbit each. Initial data reduction was
done with the standard {\sc ftools}, and the flux was normalised according
to the number of correctly functioning PCUs. For the
light curve analysis PCUs 2, 3, and 4 were used; whilst for the spectral
analysis only PCU 2 was used as the other PCUs were only turned on infrequently. 
Only the top layer of
each PCU was included in the measurements and the time resolution of
the data was 16~s. Background subtracted light curves were constructed in four 
energy bands: 2~--~4~keV, 4~--~6~keV, 6~--~10~keV and 10~--~20~keV, as well as 
a combined 2~--~10~keV band for maximum signal-to-noise. A mean X-ray 
spectrum was also extracted.

\subsection{Light curve analysis}

\begin{figure}
  \resizebox{\hsize}{!}{\includegraphics{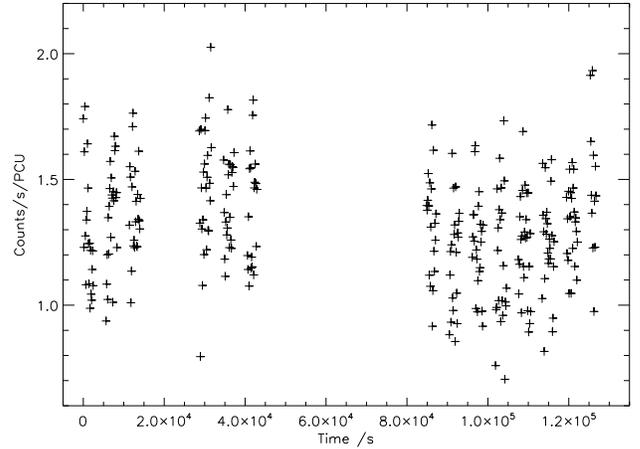}}
  \caption{2~--~10~keV background subtracted light curve of J0732.5--1331. The zero time
  corresponds to the start of the observations at JD~2454295.30035. The
  data is binned into bins of 128~s width. The typical error on each
  point is $\pm0.2$}
  \label{unfolded}
\end{figure}

\begin{figure}
  \resizebox{\hsize}{!}{\includegraphics{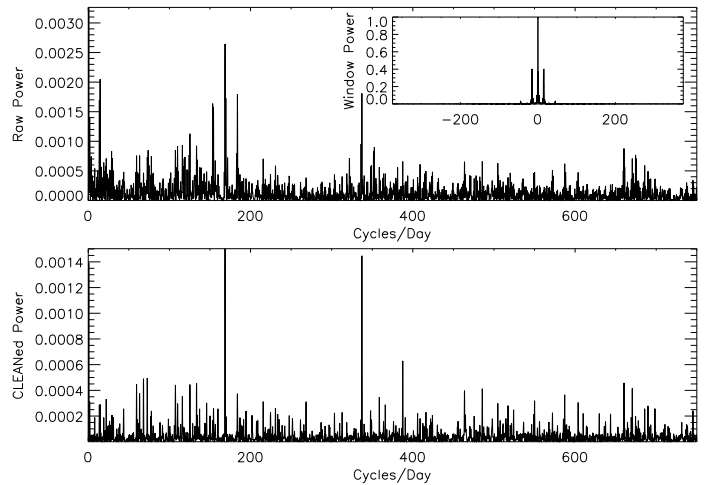}}
  \caption{{\sc clean}ed periodogram. The upper plot shows the raw periodogram,
  with the window function inset; the lower plot shows the deconvolved
  ({\sc clean}ed) periodogram.}
  \label{cleaned}
\end{figure}

The raw count rate varied between 3.9
  and 5.4~count~s$^{-1}$~PCU$^{-1}$. The background count rate, generated
  from the calibration files, varied between 2.9 and
  3.8~count~s$^{-1}$~PCU$^{-1}$. The background subtracted 2~--~10~keV light curve is shown in
Figure~\ref{unfolded}.
The data were subsequently analysed with a variable gain
  implementation of the {\sc clean}
  algorithm \citep{lehto97} 
to discover any periodicities and discount any aliasing effects. The results of 
this are shown in Figure~\ref{cleaned}.

\subsubsection{Spin Period}

\begin{figure}
  \resizebox{\hsize}{!}{\includegraphics{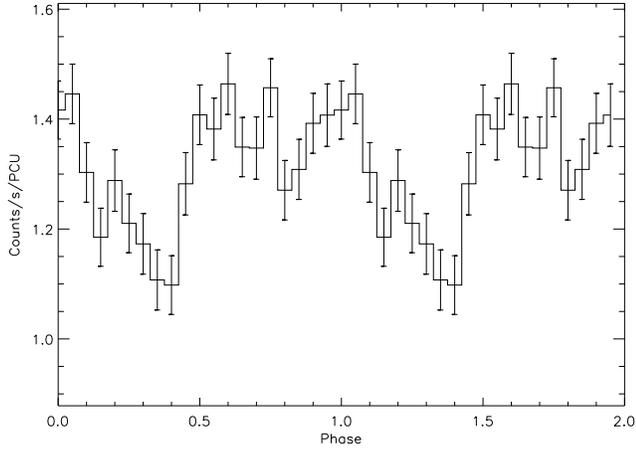}}
  \caption{2~--~10~keV light curve folded at the 512.4~s pulse period
  with an arbitrary zero point. Two cycles are shown for clarity.}
  \label{folded}
\end{figure}

Strong peaks are evident in the {\sc clean}ed periodogram at 168 cycles day$^{-1}$ 
(512~s) and at its first harmonic, in the 2~--~10~keV energy band. Similar signals are
seen in each energy band. Analysis of the peaks yields a pulsation period of 
512.4(3)~s. The data in each of the energy resolved light curves were then folded at 
the 512.4~s period, and Figure~\ref{folded} shows the result in the 2~--~10~keV energy 
band. The modulation depths of the pulse profile were then estimated by fitting a 
sinusoid to the folded data in each energy band and dividing the semi-amplitude by 
the fitted mean. The results of this are shown in Table~\ref{modulation_depths}.

\begin{table}
  \caption{Modulation depths of the pulse profile in different  energy bands. 
  Modulation depth is defined here as the semi-amplitude of a fitted sinusoid 
  divided by the fitted mean.}
  \label{modulation_depths}
  \centering
  \begin{tabular}{c c c c}
    \hline\hline
    Energy band & Modulation depth & Uncertainty & Fitted mean\\
    (keV)       & (\%)             & (\%)  & (ct~s$^{-1}$~PCU$^{-1}$)\\
    \hline
    2--10  & 8  & 1 & 1.31\\ 
    2--4   & 16 & 3 & 0.25\\
    4--6   & 7  & 2 & 0.46\\
    6--10  & 7  & 2 & 0.56\\
    10--20 & 10 & 4 & 0.28\\
    \hline
  \end{tabular}
\end{table}

\subsubsection{Orbital Period}

\begin{figure}
  \resizebox{\hsize}{!}{\includegraphics{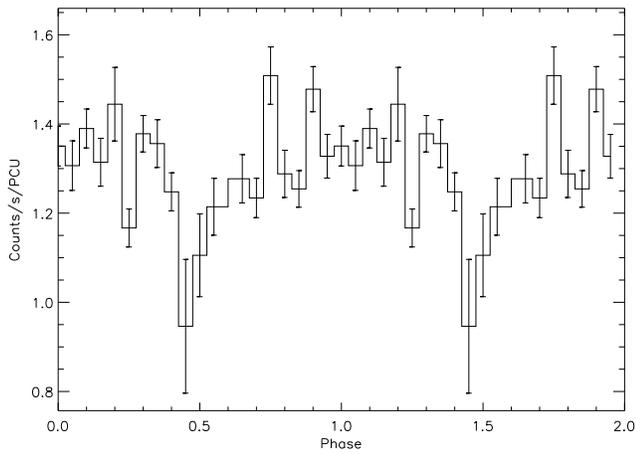}}
  \caption{2~--~10~keV light curve folded at the 11.3~h photometric period
  suggested by \cite{patterson06} with an arbitrary zero point. Two
  cycles are shown for clarity.}
  \label{folded_orbital}
\end{figure}

The windowing of the data is such that no reliable signal can be
extracted for periods of a few hours from the periodogram, therefore no
reliable orbital period can be found. Folding the X-ray data at the
5.6~h spectroscopic period \citep{thorstensen06} yields no coherent
modulation, but folding it at the 11.3~h photometric period suggested by 
\cite{patterson06} gives a plausible signal (see Figure~\ref{folded_orbital}).

\subsection{Spectral Analysis}

\begin{figure}
  \resizebox{\hsize}{!}{\rotatebox{-90}{\includegraphics{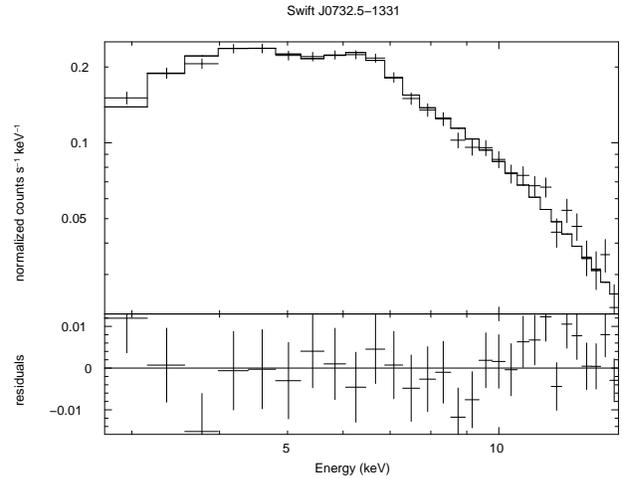}}}
  \caption{2.5~--~15 keV mean spectrum fitted with a photoelectrically
  absorbed bremsstrahlung plus iron line profile. Fitting parameters
  are $kT=37~\pm~7$~keV, $n_{\rm H}=(2.0~\pm~0.5)\times10^{22}$~cm$^{-2}$,
  iron line fitted with a Gaussian centred on 6.4~$\pm$~0.1~keV with
  $\sigma=0.3~\pm~0.1$~keV. Reduced $\chi^2$=1.0.}
  \label{spectrum}
\end{figure}

Analysis of the X-ray spectrum was carried out with the {\sc
    xspec} package. The best fit for a simple photoelectrically
    absorbed bremsstrahlung model had the parameters 
    $kT=37~\pm~7$~keV and $n_{\rm
    H}=(2.0~\pm~0.5)~\times10^{22}$~cm$^{-2}$ (reduced
    $\chi^2=3.0$). The residuals of this plot indicate the presence
    of an excess at approximately 6.5~keV. Keeping the temperature and column
density fixed and fitting a Gaussian to this region
indicates an iron line at 6.4~$\pm$~0.1~keV with a width of
$\sigma=0.3~\pm~0.1$~keV (reduced $\chi^2=1.0$), as shown in Figure~\ref{spectrum}).

\section{Discussion}

The strong X-ray signal at the 512.4(3)~s pulse period seen at all energies is 
characteristic of IPs and confirms the nature of the object. \cite{patterson06} 
found a much stronger peak at the first harmonic in their frequency analysis of 
the optical photometry data. This is characteristic of a double-peaked pulse profile 
and indicates that two emission regions can be seen. The X-ray data
reported here exhibit the same periodicity, but a somewhat different
profile. The first harmonic in the X-ray data is still present, and the 
pulse profile consequently shows a second minimum superimposed on the pulse 
maximum, but the overall profile is only marginally double-peaked. The
most likely geometry of this system is therefore such that one magnetic pole 
can always be seen, the other being behind the WD for most of the cycle. If the 
heights of the accretion columns are such that a fraction of the hidden pole's 
column can be seen at certain phases then the X-ray profile may be explained. If 
the optical emission arises from reprocessed X-rays (i.e. further up the accretion 
column) then it may be seen from both poles and this would explain the optical 
signal of \cite{patterson06}.

The modulation depth of the X-ray pulse profile is approximately constant 
above 4~keV, implying that the dominant effect shaping the profile is geometric,
probably self-occultation by the white dwarf. At the lowest energies (2~--~4~keV)
the modulation depth is higher, which  implies that phase-varying photoelectric 
absorption (as well as occultation) is the process which causes the modulation. The
spectral fitting also indicates the presence of a significant local absorbing 
column, and has parameters that are typical of other IPs.

There is still some ambiguity about the orbital period of this system. The 
chance of it being the 11.3~h period suggested by \cite{patterson06} is now
increased, given the X-ray signal seen here, but the possibility of aliasing 
in this data set means that it cannot be definitively said to be so and we cannot 
rule out the 5.6~h spectroscopic period found by \cite{thorstensen06}. The
absence of a beat period in the frequency analysis of the optical data 
does suggest that the true orbital period may be long, since no 
reprocessed radiation is seen from the face of the secondary star and 
therefore the bodies are likely to be far apart.

The lack of an X-ray beat signal in these {\em RXTE} data indicates that there
is no significant stream-fed component to the flow. This suggests a relatively
weak magnetic field strength and is consistent with the small 
$P_{\rm spin}/P_{\rm orb}$ ratio of the system \citep{norton04}, namely 
0.025 or 0.013 depending on which is the correct orbital period. 

The high temperature bremsstrahlung continuum and the presence of an iron line 
at 6.4~keV reinforce the case for J0732 to be an IP since these are both features 
often seen in other IPs (\citet{hellier04}).

Finally, we note that the average {\em RXTE} count rate (1.3~ct~s$^{-1}$~PCU$^{-1}$ in 
the 2~--~10~keV band) is 
consistent\footnote{Using webpimms: http://heasarc.gsfc.nasa.gov/Tools/w3pimms.html}
with the value obtained with the {\em Swift} satellite, indicating that the
system has not changed significantly in brightness since its discovery.

\section{Conclusion}

The unambiguous X-ray spin period detection at 512.4(3)~s, along with the 
spectral fit to an absorbed 37~keV bremsstrahlung model with an iron line,
confirm the intermediate polar status of Swift J0732.5--1331. We are 
unable to determine the orbital period from these {\em RXTE} data 
although there is some indication of modulation at the previously 
suggested photometric period of 11.3~h and none at the spectroscopic 
period of 5.6~h. To conclude we note that this system is similar, in terms of 
its small $P_{\rm spin}/P_{\rm orb}$ value, to the IPs RX~J2133.7+5107
and NY~Lup (IGR~J15479--4529). Both of these are {\em INTEGRAL} hard
X-ray sources and both also have soft X-ray components. We might
therefore expect that Swift~J0732.5--1331 would also display such
characteristics upon further study.

\bibliographystyle{aa}
\bibliography{8700}

\end{document}